\documentclass[aps,prd,preprint,groupedaddress,nofootinbib,showpacs,eqsecnum]{revtex4-1}

\usepackage{amsmath}    
\usepackage{graphicx}   

\usepackage{color}

\begin{document}


\title{Simplifying $D$-Dimensional Physical-State Sums in Gauge Theory and Gravity}


\author{Dimitrios Kosmopoulos}
\email[]{dkosmopoulos@physics.ucla.edu}
\affiliation{
Mani L. Bhaumik Institute for Theoretical Physics\\
UCLA Department of Physics and Astronomy\\
Los Angeles, CA 90095, USA\\
$\null$ \\
}


\begin{abstract}
We provide two independent systematic methods of performing
$D$-dimensional physical-state sums in gauge theory and gravity in
such a way so that spurious light-cone singularities are not
introduced.  A natural application is to generalized unitarity in the
context of dimensional regularization or theories in higher spacetime
dimensions.  Other applications include squaring matrix elements to
obtain cross sections, and decompositions in terms of gauge-invariant
tensors.
\end{abstract}



\maketitle

\tableofcontents


\newpage
\section{Introduction}

The past years have seen remarkable advances to our ability to
calculate scattering amplitudes in perturbative quantum field
theory. On the one hand, much of this progress relies on choices of
variables that exploit the four-dimensional nature of the kinematics,
such as spinor-helicity~\cite{SpinorHelicity} or
momentum-twistor~\cite{Hodges:2009hk} variables. On the other hand,
for certain problems it is favorable to work in arbitrary dimension
$D$. For example, $D$-dimensional methods proved useful in 
the recent evaluation of the conservative two-body
Hamiltonian for spinless black holes to order $G^3$~\cite{3PM},
relevant to gravitational-wave physics studied by the LIGO and Virgo
collaborations~\cite{gravWaveDiscovery}.

In multiloop calculations, the preferred regularization scheme is
dimensional regularization~\cite{Collins}. Occasionally, subtleties
arise when one combines four-dimensional methods with dimensional
regularization. In these instances $D$-dimensional methods are
necessary, as was the case for example in the recent reexamination of
the two-loop counterterm of pure
gravity~\cite{Bern:2015xsa}. Furthermore, we are often interested in
performing a calculation in a generic dimension. The calculation of
the gravitational potential between two scalar particles in arbitrary
dimension at order $G^2$~\cite{Cristofoli:2020uzm} is a recent illustration.

A prominent method used in $D$-dimensional calculations is generalized unitarity. Generalized unitarity was orginally developed for four-dimensional computations~\cite{Unitarity}, but has since been extended to higher dimensions~\cite{dDimUnitarity}.  $D$-dimensional generalized unitarity has been employed in calculations pertaining to phenomenology at the Large Hadron Collider (LHC) (see for e.g. ~\cite{Ellis:2008qc,Badger:2017jhb}), as well as in the study of supersymmetric theories~\cite{Gehrmann:2011xn}. It meshes well with other modern amplitudes techniques, such as the double copy~\cite{Kawai:1985xq,BCJ}, and as such is a natural tool for computations in gravitational theories~\cite{Carrasco:2013ypa,Bern:2015ooa,Mogull:2015adi}. Recently,  Ref.~\cite{Engelund:2013fja} used generalized unitarity for a worldsheet theory. These important and diverse results underline the significance of simplifying as much as possible the implementation of $D$-dimensional generalized unitarity.
 
A difficulty we encounter when we work in $D$ dimensions is that physical-state sums
for gluons and gravitons introduce spurious light-cone singularities that
complicate the calculation. 
If we
do not eliminate these spurious singularities at the level of the
integrand, we have to regularize them with the
Mandelstam-Leibbrandt~\cite{Mandelstam-Leibbrandt} or
principle-value~\cite{PVprescription} prescription for example, which
complicates integration.

In this paper we develop two methods for performing these sums
so that we do not introduce spurious singularities.
Refs.~\cite{diVecchia,3PM,Bern:2020buy} showed that in certain four-point 
amplitudes by appropriate arrangements the spurious light-cone
singularities automatically drop out in generalized-unitary cuts.
Here we provide methods to systematically eliminate
such spurious singularities from any generalized-unitarity
cut or any sewing involving gauge-invariant quantities at any loop order.

\begin{figure}[t]
\includegraphics[scale=.6]{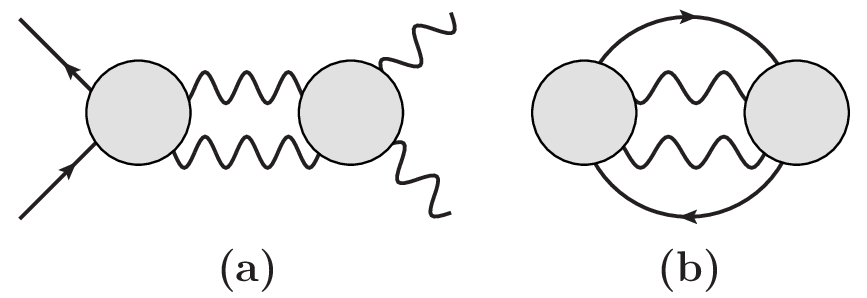}
\caption{ \small Examples of calculations
       where we may apply the techniques of this paper:
       (a) integrand-level generalized-unitarity cut and (b)
       squared matrix elements for cross sections.  The blobs
       represent amplitudes. All exposed lines are taken as on
       shell. The internal exposed lines indicate gauge-particle legs
       that we intend to sew. }
\label{SewingExamples}
\end{figure}

We may apply our methods to a variety of situations, some
of which we depict in Fig.~\ref{SewingExamples}. For calculations
based on generalized unitarity~\cite{Unitarity}, while in some cases
it is sufficient to compute the generalized-unitarity cuts in
four-dimensions \cite{4DimUnitarity}, in others we need to know them
in $D=4-2\epsilon$ dimensions \cite{dDimUnitarity}. In some cases, matrix-element squares,
useful in calculations of cross sections, are calculated in $D$
dimensions \cite{scatteringCrossSections}. Furthermore, a useful technique
that relies on physical-state sums is the decomposition of an
amplitude into gauge-invariant tensors \cite{gaugeInvariantTensors}.

In this paper we provide two independent methods of performing the
physical-state sums so that we do not introduce spurious singularities. In the first method we identify gauge-invariant
subpieces (two in gauge theory and three in gravity) of our expression
and perform the physical-state sum for each subpiece separately. This
allows us to substitute the physical-state projectors with
replacements rules for the individual subpieces that do not contain
spurious singularities. In our second method we provide a
simple systematic way to make a gauge-invariant quantity obey the
generalized Ward identity, such that the spurious-singularity pieces automatically drop out of the physical-state sum. We find that there are
certain limitations in the applicability of the second approach, which
however are not relevant for many practical purposes.

We focus on theories that contain scalars, photons, gluons and gravitons for concreteness. However, we may straightforwardly apply our techniques to theories with different matter content. Some examples are theories that contain fermions or higher-spin fields~\cite{Bern:2020buy,Singh:1974qz}.

We organize the remainder of the paper as follows. In Sect.~\ref{backAndDef} we review a few properties that are useful for our purposes and establish our conventions. In Sect.~\ref{spFreeRules} we discuss our first method of performing the physical-state sums. We describe how to isolate the gauge-invariant subpieces of an expression and how to perform the physical-state sum for each one. In Sect.~\ref{GWIsec} we describe our second method. We explain how to bring an expression in the appropriate form and how to use that form to remove the spurious-singularity pieces from the physical-state projectors. We present our conclusions in Sect.~\ref{ConclusionsSec}.

\section{Background and definitions\label{backAndDef}}

In this paper we study gauge and gravitational theories. We refer to the particles associated with these theories, namely the gluon and the graviton, as gauge particles. We describe the state of a gluon by a null polarization vector $\varepsilon^\mu$, as appropriate for circular polarization. We express the polarization tensor that describes the state of the graviton as a product of two factors of the polarization vector of the gluon, $\varepsilon^{\mu \nu} = \varepsilon^\mu \varepsilon^\nu$. This polarization tensor is traceless since the gluon polarization vectors are taken to be null. With this construction we may collectively describe the gauge particle by its polarization vector $\varepsilon^\mu$.

We analyse the sewing of gauge-invariant quantities. By sewing we refer to performing the physical-state sum of some gauge-particle legs that belong to these quantities. We denote gauge-invariant quantities by $\mathcal{A}$ and refer to them as amplitudes. Our results, however, apply to any gauge-invariant quantities. They apply for example to higher-loop generalized-unitarity cuts~\cite{Unitarity} or gauge-invariant tensors~\cite{gaugeInvariantTensors}.

When a quantity is gauge invariant, then it satisfies the Ward identity (WI) for each gauge particle. The WI states that when the polarization vector of a single gauge particle $\varepsilon$ is replaced by the particle's momentum $p$, then the amplitude vanishes:
\begin{equation}
\mathcal{A}\big|_{\varepsilon_\mu \rightarrow p_\mu} = 0 \, \quad \text{(gauge theory)} \,, \quad 
\mathcal{A}\big|_{\varepsilon_\mu \varepsilon_\nu \rightarrow p_\mu q_\nu} = 0 \, \quad \text{(gravity)} .
\label{WI}
\end{equation}
In the gravitational case, one of the two factors of the polarization vector is replaced by the corresponding momentum, while the other one by an arbitrary vector $q^\nu$\footnote{In practice, it is convenient to formally distinguish the two factors (eg. $\varepsilon^{\mu \nu} = \varepsilon^\mu \tilde{\varepsilon}^\nu$) and only set them equal ($\tilde{\varepsilon}^\mu=\varepsilon^\mu$) at the end of the calculation.}. 
We emphasize that this property is only true upon use of the on-shell conditions. In gravity we have to use the on-shell conditions both before and after the above replacement. Namely, one must use the mass-shell condition for every particle ($p_i^2 = m_i^2$), momentum conservation ($\sum_i p_i = 0$, where we take all particles to be outgoing), and the fact that polarization vectors are null ($\varepsilon_i^2 = 0$) and transverse ($\varepsilon_i \cdot p_i = 0$).

When sewing gauge-invariant quantities, one is instructed to sum over the physical polarizations of the gauge particles of interest. This sum is equal to the physical-state projector. We occasionally refer to performing this sum as inserting the projector. In gauge theory we have
\begin{equation}
P^{\mu \nu} (p,q) = \sum_{\text{pols.}} \varepsilon^\mu (-p) \varepsilon^\nu (p) = \eta^{\mu \nu} - \frac{q^\mu p^\nu + p^\mu q^\nu}{q\cdot p},
\label{prYM}
\end{equation}
while in gravity
\begin{equation}
P^{\mu \nu \alpha \beta} (p,q) = \sum_{\text{pols.}} \varepsilon^{\mu \nu} (-p) \varepsilon^{\alpha \beta} (p) = \frac{1}{2} \left( 
	P^{\mu \alpha} P^{\nu \beta} + P^{\nu \alpha} P^{\mu \beta} \right) - \frac{1}{D-2} P^{\mu \nu} P^{\alpha \beta} .
\label{prGR}
\end{equation}
In both cases $p$ is the momentum of the gauge particle in question, $q$ is an arbitrary null vector and $D$ is the dimension of spacetime. We observe that insertions of the physical-state projectors introduce spurious light-cone singularities $1/{q\cdot p}$. We call them spurious because they drop out from the final expression, appearing only in intermediate steps of calculations.

In this paper we develop strategies that allow us to effectively\footnote{We use the word effectively because the replacement might also need an overall factor. We discuss the details in the following sections.} make the following replacement in Eqs.~(\ref{prYM}) and (\ref{prGR}):
\begin{equation}
P^{\mu \nu} (p,q)  \rightarrow \eta^{\mu \nu}. 
\label{simplify}
\end{equation}
Specifically, for gravity we effectively get
\begin{equation}
P^{\mu \nu \alpha \beta} (p,q) \rightarrow \frac{1}{2}\left(\eta_{\mu \alpha} \eta_{\nu \beta} + 
	\eta_{\mu \beta} \eta_{\nu \alpha} - \frac{2}{D-2}\eta_{\mu \nu}\eta_{\alpha \beta} \right),
\end{equation}
where on the right-hand side we recognize the de Donder projector. 
Our methods allow us to use Feynman-gauge-like and de Donder-gauge-like sewing rules, while at the same time not introduce any ghost degrees of freedom. We do not achieve this by choosing a specific gauge. Rather, we phrase the whole discussion in terms of on-shell objects.

\section{Spurious-singularity-free replacement rules\label{spFreeRules}}

In this section we derive a set of replacement rules that may be used to perform the physical-state sums in gauge theory and gravity. These rules do not contain spurious singularities while using them is equivalent to using Eqs.~(\ref{prYM}) and (\ref{prGR}). The only requirement for applying these rules is that the quantities being sewn obey the WI.

As a warm-up to the general case, we consider a simple example in scalar QED that demonstrates the basic idea. We proceed to discuss our method in general. Then, we provide a summary of our results. Finally, we conclude this section by studying an involved example in detail. 

\subsection{Demonstration in a simple example\label{demoRulesSec}}

\begin{figure}[t]
    \centering
	\includegraphics[scale=0.5]{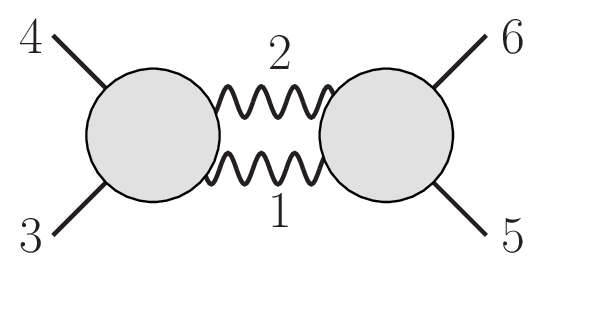}
	\caption{\label{sQEDcutFig}The scalar-QED generalized-unitarity cut studied in Sect.~\ref{demoRulesSec}. The two blobs are Compton amplitudes in this theory. Solid lines correspond to scalar particles and wiggly lines correspond to photons. External momenta are taken outgoing while internal momenta flow to the right. All exposed lines are taken as on shell.}
\end{figure} 

Here we wish to demonstrate our method with a simple example. In particular, we compute the one-loop generalized-unitarity cut of Fig.~\ref{sQEDcutFig} without introducing spurious singularities. 
We take the external particles to be massive scalars while the internal particles are photons. 
We compute the generalized-unitarity cut in question by sewing two factors of the Compton amplitude of scalar QED.
At the level of the Compton amplitudes, all particles are taken as on shell.
We choose to consider scalar QED due to the compactness of the expression. 
Refs.~\cite{diVecchia,Bern:2020buy} calculated corresponding cuts in gravity. 

\begin{figure}[t]
    \includegraphics[scale=.8]{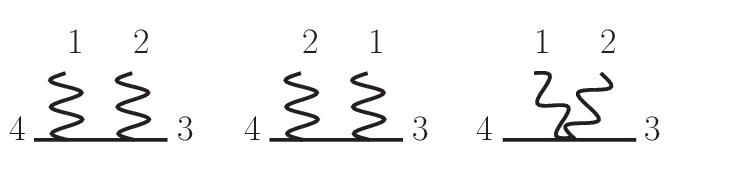}
        \caption{\label{sQEDFig}\small The three Feynman diagrams we need to calculate in order to get the scalar-QED amplitudes of Eq.~(\ref{sQEDlrAmpsEq}). The solid line represents a scalar particle while the wiggly lines represent photons. We take all momenta to be outgoing.}        
\end{figure}

We set the couplings to unity for convenience.  We denote the two Compton amplitudes that enter our example by $\mathcal{A}_L$ and $\mathcal{A}_R$. We have
\begin{align}
\begin{split}
\mathcal{A}_L &= 2 i \left( \frac{p_4 \cdot \varepsilon_1 p_3 \cdot \varepsilon_2}{p_1 \cdot p_4} + 
\frac{p_4 \cdot \varepsilon_2 p_3 \cdot \varepsilon_1}{p_1 \cdot p_3} + \varepsilon_1 \cdot \varepsilon_2 \right), \\
\mathcal{A}_R &= 2 i \left( \frac{p_6 \cdot \varepsilon_{-1} p_5 \cdot \varepsilon_{-2}}{-p_1 \cdot p_6} + 
\frac{p_6 \cdot \varepsilon_{-2} p_5 \cdot \varepsilon_{-1}}{-p_1 \cdot p_5} + \varepsilon_{-1} \cdot \varepsilon_{-2} \right),
\label{sQEDlrAmpsEq}
\end{split}
\end{align}
where $\varepsilon_{i} \equiv \varepsilon(p_i)$ and $\varepsilon_{-i} \equiv \varepsilon(-p_i)$. We present the Feynman diagrams needed to calculate $\mathcal{A}_L$ in Fig.~\ref{sQEDFig}.
We use the on-shell conditions to reduce $\mathcal{A}_L$ to a basis of Lorentz-invariant products. Namely, we solve momentum conservation as $p_2 = -p_1-p_3-p_4$ and impose $\varepsilon_2 \cdot p_1 = -\varepsilon_2 \cdot p_3 - \varepsilon_2 \cdot p_4$ and $p_3 \cdot p_4 = -p_1 \cdot p_3 - p_1 \cdot p_4 - m_L^2$, where $m_L$ is the mass of the scalar.  We obtain $\mathcal{A}_R$ by the appropriate relabelling.

We wish to sew both photon-legs 1 and 2, so that we obtain the generalized-unitarity cut depicted in Fig.~\ref{sQEDcutFig}.  We start by summing  over the physical polarizations of photon-leg 1. We denote this sum as $\sum_1$. We have
\begin{equation}
\sum_1 \mathcal{A}_L \mathcal{A}_R = \mathcal{C}_0^{\mu \nu} \sum_1 \varepsilon_{1 \mu} \varepsilon_{-1 \nu},
\end{equation}
where 
\begin{equation}
\mathcal{C}_0^{\mu \nu} \equiv 
-4 \left( \frac{p_3 \cdot \varepsilon_2}{p_1 \cdot p_4} p_4^\mu + 
\frac{p_4 \cdot \varepsilon_2}{p_1 \cdot p_3} p_3^\mu + \varepsilon_2^\mu \right)
\left( \frac{p_5 \cdot \varepsilon_{-2}}{-p_1 \cdot p_6} p_6^\nu + 
\frac{p_6 \cdot \varepsilon_{-2}}{-p_1 \cdot p_5} p_5^\nu + \varepsilon_{-2}^\nu \right).
\end{equation}
We build $\mathcal{C}_0^{\mu \nu}$ out of the remaining momenta and polarization vectors. Importantly, it does not contain the metric $\eta^{\mu \nu}$. Observe that 
\begin{equation}
\mathcal{C}_0^{\mu \nu} p_{1 \mu} = \mathcal{C}_0^{\mu \nu} p_{1 \nu} = 0.
\label{c0Transversality}
\end{equation}
This is a consequence of the WI for particle 1, obeyed by $\mathcal{A}_L$ and $\mathcal{A}_R$. We can see this explicitly as follows. If we define $\mathcal{A}_L^\mu$ by $\mathcal{A}_L = \varepsilon_{1 \mu} \mathcal{A}_L^\mu$ and similarly for $\mathcal{A}_R^\nu$, then we have 
\begin{equation}
\mathcal{C}_0^{\mu \nu} = \mathcal{A}_L^\mu \mathcal{A}_R^\nu.
\end{equation}
Then the WI for the two amplitudes (Eq.~(\ref{WI})) reads
\begin{equation}
p_{1 \mu} \mathcal{A}_L^\mu = p_{1 \nu} \mathcal{A}_R^\nu = 0.
\end{equation}
We call $\mathcal{A}_L^\mu$ and $\mathcal{A}_R^\nu$ transverse because they obey the above property. We observe that in this case we may write $\mathcal{C}_0^{\mu \nu}$ as a product of two transverse objects.

Using Eq.~(\ref{c0Transversality}) we may simplify the insertion of the projector. Focusing on the spurious singularity piece of the projector (Eq.~(\ref{prYM})), we have
\begin{equation}
\left( \sum_1 \mathcal{A}_L \mathcal{A}_R \right) _\text{spurious sing.} = 
\frac{1}{p_1 \cdot q} \left(  \mathcal{C}_0^{\mu \nu} p_{1 \mu} q_{\nu} + \mathcal{C}_0^{\mu \nu} p_{1 \nu} q_{\mu} \right) = 0.
\end{equation}
Then the full expression becomes
\begin{align}
\begin{split}
\sum_1 \mathcal{A}_L \mathcal{A}_R &= \mathcal{C}_0^{\mu \nu} \sum_1 \varepsilon_{1 \mu} \varepsilon_{-1 \nu} 
= \mathcal{C}_0^{\mu \nu} \eta_{\mu \nu} \\
&= 4 \Bigg( \frac{p_4 \cdot p_6 p_3 \cdot \varepsilon_2 p_5 \cdot \varepsilon_{-2}}{p_1 \cdot p_4 p_1 \cdot p_6} + 
\frac{p_4 \cdot p_5 p_3 \cdot \varepsilon_2 p_6 \cdot \varepsilon_{-2}}{p_1 \cdot p_4 p_1 \cdot p_5} -
\frac{p_3 \cdot \varepsilon_2 p_4 \cdot \varepsilon_{-2} }{p_1 \cdot p_4 } \\
&+ 
\frac{p_3 \cdot p_6 p_4 \cdot \varepsilon_2 p_5 \cdot \varepsilon_{-2}}{p_1 \cdot p_3 p_1 \cdot p_6} + 
\frac{p_3 \cdot p_5 p_4 \cdot \varepsilon_2 p_6 \cdot \varepsilon_{-2}}{p_1 \cdot p_3 p_1 \cdot p_5} -
\frac{p_4 \cdot \varepsilon_2 p_3 \cdot \varepsilon_{-2} }{p_1 \cdot p_3 } \\
&+
\frac{p_6 \cdot \varepsilon_2 p_5 \cdot \varepsilon_{-2}}{p_1 \cdot p_6} + 
\frac{p_5 \cdot \varepsilon_2 p_6 \cdot \varepsilon_{-2}}{p_1 \cdot p_5} -
\varepsilon_2 \cdot \varepsilon_{-2}
\Bigg).
\end{split}
\end{align}

Next, we wish to sew photon-leg 2. Before we proceed, we reduce our expression to a basis of Lorentz-invariant products.
Namely, we solve momentum conservation as $p_1 = -p_2+p_5+p_6$ and $p_3 = -p_4-p_5-p_6$. Any other choice would be equally valid. We impose the trasversality conditions, $\varepsilon_{\pm 2} \cdot p_2 = 0$, and the mass-shell conditions. We find 
\begin{align}
\begin{split}
\sum_1 \mathcal{A}_L \mathcal{A}_R &= \mathcal{C}_0^{\mu \nu} \eta_{\mu \nu} \\
&= 4 \Bigg( -\frac{ (p_2 \cdot p_5 + p_2 \cdot p_6 + p_4 \cdot p_5) (p_4+p_5+p_6) \cdot \varepsilon_2 \, p_5 \cdot \varepsilon_{-2}}{p_2 \cdot p_5 \, p_2 \cdot (p_4 + p_5 + p_6)} \\ &+ 
\frac{p_4 \cdot p_5 (p_4+p_5+p_6) \cdot \varepsilon_2 \, p_6 \cdot \varepsilon_{-2}}{p_2 \cdot p_6 p_2 \cdot (p_4+p_5+p_6)} 
- \frac{(p_4+p_5+p_6) \cdot \varepsilon_2 \, p_4 \cdot \varepsilon_{-2} }{p_2 \cdot (p_4+p_5+p_6) } \\
&+ 
\frac{p_4 \cdot p_5 p_4 \cdot \varepsilon_2 p_5 \cdot \varepsilon_{-2}}{p_2 \cdot p_4 p_2 \cdot p_5} - 
\frac{( p_2 \cdot p_5 + p_2 \cdot p_6 + p_4 \cdot p_5 ) p_4 \cdot \varepsilon_2 p_6 \cdot \varepsilon_{-2}}{p_2 \cdot p_4 p_2 \cdot p_6} \\ 
&+ \frac{p_4 \cdot \varepsilon_2 (p_4+p_5+p_6) \cdot \varepsilon_{-2} }{p_2 \cdot p_4 }
+ \frac{p_6 \cdot \varepsilon_2 p_5 \cdot \varepsilon_{-2}}{p_2 \cdot p_5} + 
\frac{p_5 \cdot \varepsilon_2 p_6 \cdot \varepsilon_{-2}}{p_2 \cdot p_6} -
\varepsilon_2 \cdot \varepsilon_{-2}
\Bigg).
\end{split}
\end{align}

Now we may sew particle 2,
\begin{equation}
\sum_{1,2} \mathcal{A}_L \mathcal{A}_R = 
 \sum_2 \mathcal{C}_0^{\mu \nu} \eta_{\mu \nu} = \tilde{\mathcal{C}}_0^{\mu \nu} \sum_2 \varepsilon_{2 \mu} \varepsilon_{-2 \nu} -4 \sum_2 \varepsilon_2 \cdot \varepsilon_{-2},
\end{equation}
where 
\begin{align}
\begin{split}
\tilde{\mathcal{C}}_0^{\mu \nu} &=
4 \Bigg( -\frac{ (p_2 \cdot p_5 + p_2 \cdot p_6 + p_4 \cdot p_5) (p_4^\mu+p_5^\mu+p_6^\mu) \, p_5^\nu }{p_2 \cdot p_5 \, p_2 \cdot (p_4 + p_5 + p_6)} \\ &+ 
\frac{p_4 \cdot p_5 (p_4^\mu+p_5^\mu+p_6^\mu) \, p_6^\nu }{p_2 \cdot p_6 p_2 \cdot (p_4+p_5+p_6)} 
- \frac{(p_4^\mu+p_5^\mu+p_6^\mu) \, p_4^\nu }{p_2 \cdot (p_4+p_5+p_6) } \\
&+ 
\frac{p_4 \cdot p_5 p_4^\mu p_5^\nu }{p_2 \cdot p_4 p_2 \cdot p_5} - 
\frac{( p_2 \cdot p_5 + p_2 \cdot p_6 + p_4 \cdot p_5 ) p_4^\mu p_6^\nu }{p_2 \cdot p_4 p_2 \cdot p_6} \\ 
&+ \frac{p_4^\mu (p_4^\nu+p_5^\nu+p_6^\nu) }{p_2 \cdot p_4 }
+ \frac{p_6^\mu p_5^\nu }{p_2 \cdot p_5} + 
\frac{p_5^\mu p_6^\nu }{p_2 \cdot p_6}
\Bigg).
\end{split}
\end{align}
Again, $\tilde{\mathcal{C}}_0^{\mu \nu}$ is built out of the remaining momenta and does not contain the metric. One may verify that 
\begin{equation}
\tilde{\mathcal{C}}_0^{\mu \nu} p_{2 \mu} = \tilde{\mathcal{C}}_0^{\mu \nu} p_{2 \nu} = 0.
\end{equation} 
This time this property does not straightforwardly follow from the WI of the two amplitudes (indeed, we cannot write $\tilde{\mathcal{C}}_0^{\mu \nu}$ as a product of two transverse objects). We discuss this property in detail in the next subsection. A hint about why it holds is that the whole expression obeys the WI for particle 2 and so does the term $\varepsilon_2 \cdot \varepsilon_{-2}$ by itself, since $\varepsilon_{\pm 2} \cdot p_2 =0$.

Delaying its explanation, we may use the above observation to simplify the sewing of particle 2:
\begin{equation}
\sum_{1,2} \mathcal{A}_L \mathcal{A}_R = 
 \tilde{\mathcal{C}}_0^{\mu \nu} \sum_2 \varepsilon_{2 \mu} \varepsilon_{-2 \nu} -4 \sum_2 \varepsilon_2 \cdot \varepsilon_{-2} = \tilde{\mathcal{C}}_0^{\mu \nu} \eta_{\mu \nu} -4 \sum_2 \varepsilon_2 \cdot \varepsilon_{-2}.
\end{equation}
We complete our example by performing the remaining sum, 
\begin{equation}
\sum_2 \varepsilon_2 \cdot \varepsilon_{-2} = D-2,
\end{equation}
where $D$ is the dimension of spacetime. We obtain the above by dotting Eq.~(\ref{prYM}) with the metric $\eta_{\mu \nu}$. Finally,
\begin{align}
\begin{split}
\sum_{1,2} \mathcal{A}_L \mathcal{A}_R &= \tilde{\mathcal{C}}_0^{\mu \nu} \eta_{\mu \nu} -4 (D-2) \\
&= 
4 \Bigg( -\frac{ (p_2 \cdot p_5 + p_2 \cdot p_6 + p_4 \cdot p_5) (p_4+p_5+p_6) \cdot p_5 }{p_2 \cdot p_5 \, p_2 \cdot (p_4 + p_5 + p_6)} \\ &+ 
\frac{p_4 \cdot p_5 (p_4+p_5+p_6) \cdot p_6 }{p_2 \cdot p_6 p_2 \cdot (p_4+p_5+p_6)} 
- \frac{(p_4+p_5+p_6) \cdot p_4 }{p_2 \cdot (p_4+p_5+p_6) } \\
&+ 
\frac{p_4 \cdot p_5 p_4 \cdot p_5 }{p_2 \cdot p_4 p_2 \cdot p_5} - 
\frac{( p_2 \cdot p_5 + p_2 \cdot p_6 + p_4 \cdot p_5 ) p_4 \cdot p_6 }{p_2 \cdot p_4 p_2 \cdot p_6} \\ 
&+ \frac{p_4 \cdot (p_4+p_5+p_6) }{p_2 \cdot p_4 }
+ \frac{p_6 \cdot p_5 }{p_2 \cdot p_5} + 
\frac{p_5 \cdot p_6 }{p_2 \cdot p_6}
\Bigg) 
-4 (D-2).
\end{split}
\end{align}
We conclude that we may perform both sewings without introducing any spurious singularities.

\subsection{The general case}

Now we take up the discussion of the general case of summing over the physical polarizations of a gauge particle.  We denote the polarization vectors and momenta of the gauge-particle legs being sewn as $\varepsilon_+ \equiv \varepsilon(p)$ and $p$, and $\varepsilon_- \equiv \varepsilon(-p)$ and $(-p)$ respectively.  We write $\sum_\textrm{pols.}$ for the sum over the physical polarizations of the gauge particle.

\textit{Claim:} Given a quantity $C$ that obeys the Ward identity (Eq.~(\ref{WI})) for both $\varepsilon_+$ and $\varepsilon_-$, we may perform the sum
\begin{equation}
\sum_\textrm{pols.} C,
\end{equation}
in terms of a set of spurious-singularity-free replacements rules. We collect these rules in Eq.~(\ref{ymRules}) for gauge theory and Eq.~(\ref{grRules}) for gravity.

We proceed to prove the above claim in the cases of gauge theory and gravity separately.

\subsubsection{Proof in gauge theory}

In the case of gauge theory, the quantity $C$ may be written as
\begin{equation}
C = \mathcal{C}_0^{\mu \nu} \varepsilon_{+\mu} \varepsilon_{-\nu} + \mathcal{C}_1  \varepsilon_+ \cdot \varepsilon_- ,
\end{equation}
where we build $\mathcal{C}_0^{\mu \nu}$ and $\mathcal{C}_1$ out of the remaining momenta and polarization vectors. We note that $\mathcal{C}_0^{\mu \nu}$ does not contain the metric $\eta^{\mu \nu}$. In other words, there is no ambiguity in splitting $C$ into these two pieces.

At this point we reduce $C$ to a basis of Lorentz-invariant products using momentum conservation and the on-shell conditions. Specifically, we impose the transversality condition ($\varepsilon_{\pm} \cdot p =0$). For example, if we choose to solve momentum conservation such that $p$ is one of the independent momenta appearing in our basis, then we may decompose $\mathcal{C}_0^{\mu \nu}$ as
\begin{equation}
\mathcal{C}_0^{\mu \nu} = Q^{\mu \nu} + Q_+^{\mu} p^\nu + Q_-^{\nu} p^\mu + Q_0 p^\mu p^\nu.
\end{equation}
Then, due to the transversality condition we have
\begin{equation}
\mathcal{C}_0^{\mu \nu} \varepsilon_{+\mu} \varepsilon_{-\nu} = Q^{\mu \nu} \varepsilon_{+\mu} \varepsilon_{-\nu}.
\end{equation}
We note that we do not have to choose the above basis.

Next, being gauge invariant, $C$ obeys the WI for the two legs in question:
\begin{equation}
C \vert_{\varepsilon_+ \rightarrow p} = 0, \quad \quad C \vert_{\varepsilon_- \rightarrow p} = 0.
\end{equation}
We observe that so does $\varepsilon_+ \cdot \varepsilon_-$ since $\varepsilon_{\pm} \cdot p =0$. We conclude that the last term must also obey the WI:
\begin{equation}
\mathcal{C}_0^{\mu \nu} \varepsilon_{+\mu} p_{\nu} = 0, \quad \quad \mathcal{C}_0^{\mu \nu} p_{\mu} \varepsilon_{-\nu} = 0.
\label{c0WI}
\end{equation}
We now argue that the stronger condition 
\begin{equation}
\mathcal{C}_0^{\mu \nu} p_{\nu} = 0, \quad \quad \mathcal{C}_0^{\mu \nu} p_{\mu}  = 0,
\end{equation}
follows from the above. Indeed, that would not be the case only if we needed to use some of the special properties of the polarization vectors in Eq.~(\ref{c0WI}). Those properties are 
\begin{equation}
\varepsilon_\pm \cdot p =0 \quad \text{and} \quad \varepsilon_\pm^2 = 0.
\end{equation}
The null condition is not required, since $\mathcal{C}_0^{\mu \nu}$ does not contain $\varepsilon_\pm$. The transversality condition is also not required, since we have reduced $\mathcal{C}_0^{\mu \nu}$ to a basis of Lorentz-invariant products (this is most easily seen in a basis where $p$ is one of the independent momenta).

Using this property we may cancel the spurious-singularity piece of the projector
\begin{equation}
\left( \mathcal{C}_0^{\mu \nu} \sum_\textrm{pols.} \varepsilon_{+\mu} \varepsilon_{-\nu} \right)  _\text{spurious sing.} = \frac{1}{p \cdot q} \left( \mathcal{C}_0^{\mu \nu} p_{\mu} q_{\nu} + \mathcal{C}_0^{\mu \nu} p_{\nu} q_{\mu} \right) = 0.
\end{equation}
Then we find
\begin{equation}
\sum_\textrm{pols.} C = \mathcal{C}_0^{\mu \nu} \sum_\textrm{pols.} \varepsilon_{+\mu} \varepsilon_{-\nu}
+ \mathcal{C}_1 \sum_\textrm{pols.}  \varepsilon_+ \cdot \varepsilon_- = 
\mathcal{C}_0^{\mu \nu} \eta_{\mu \nu} + \mathcal{C}_1 (D-2),
\label{GaugeRulesResult}
\end{equation}
where 
\begin{equation}
\sum_\text{pols.} \varepsilon_+ \cdot \varepsilon_- = D-2,
\label{GaugeExactProduct}
\end{equation}
and $D$ is the spacetime dimension.

From Eq.~(\ref{GaugeRulesResult}) we see that performing the physical-state sum at hand is equivalent to using the replacement rule
\begin{equation}
x_1^\mu x_2^\nu \sum_\textrm{pols.} \varepsilon_{+\mu} \varepsilon_{-\nu} \rightarrow x_1^\mu x_2^\nu \eta_{\mu \nu},
\end{equation}
supplemented by Eq.~(\ref{GaugeExactProduct}) in the individual terms in $\sum_\textrm{pols.}C$. By $x_i$ we refer to the various vectors that appear in the problem.
In this way, by treating the term $\varepsilon_+ \cdot \varepsilon_-$ separately we manage to sum over the physical polarizations of the gauge particle without introducing any spurious singularities.

\subsubsection{Proof in gravity}

The case of gravity follows in a similar manner. We may write
\begin{equation}
C = 
\mathcal{C}_0^{\mu \nu \alpha \beta} \varepsilon_{+\mu} \varepsilon_{+\nu} \varepsilon_{-\alpha} \varepsilon_{-\beta}
+ \mathcal{C}_1^{\mu \nu} \varepsilon_{+\mu} \varepsilon_{-\nu} (\varepsilon_+ \cdot \varepsilon_-)
+ \mathcal{C}_2 (\varepsilon_+ \cdot \varepsilon_-)^2 ,
\end{equation}
where we build $\mathcal{C}_0^{\mu \nu \alpha \beta}$, $\mathcal{C}_1^{\mu \nu}$ and $\mathcal{C}_2$ out of the remaining momenta and polarization vectors. This splitting is unique, since $\mathcal{C}_0^{\mu \nu \alpha \beta}$ and $\mathcal{C}_1^{\mu \nu}$ do not contain any terms proportional to the metric or combinations of the metric. In writing the above we make use of the fact that the polarization vectors are null $(\varepsilon_\pm^2 =0)$. As before, we must also impose the transversality condition $(\varepsilon_\pm \cdot p =0)$.

We use the fact that $C$ and $\varepsilon_+ \cdot \varepsilon_-$ obey the WI to deduce that all three terms individually obey the WI. Then, we may show that
\begin{equation}
\mathcal{C}_0^{\mu \nu \alpha \beta} p_\mu = \mathcal{C}_0^{\mu \nu \alpha \beta} p_\nu = \mathcal{C}_0^{\mu \nu \alpha \beta} p_\alpha = 
\mathcal{C}_0^{\mu \nu \alpha \beta} p_\beta = 0,
\end{equation}
and
\begin{equation}
\mathcal{C}_1^{\mu \nu} p_\mu = \mathcal{C}_1^{\mu \nu} p_\nu = 0.
\end{equation}
It then follows from Eq.~(\ref{prGR}) that
\begin{equation}
\mathcal{C}_0^{\mu \nu \alpha \beta} \sum_\textrm{pols.} \varepsilon_{+\mu} \varepsilon_{+\nu} \varepsilon_{-\alpha} \varepsilon_{-\beta} = \mathcal{C}_0^{\mu \nu \alpha \beta} \frac{1}{2}\left(\eta_{\mu \alpha} \eta_{\nu \beta} + 
	\eta_{\mu \beta} \eta_{\nu \alpha} - \frac{2}{D-2}\eta_{\mu \nu}\eta_{\alpha \beta} \right),
\end{equation}
and
\begin{equation}
\mathcal{C}_1^{\mu \nu} \sum_\textrm{pols.} 
	\varepsilon_{+\mu} \varepsilon_{-\nu} (\varepsilon_+ \cdot \varepsilon_-) =  
	\mathcal{C}_1^{\mu \nu} \eta_{\mu \nu} \frac{D(D-3)/2}{D-2}.
\end{equation}
Finally, for $(\varepsilon_+ \cdot \varepsilon_-)^2$ we get
\begin{equation}
\sum_\text{pols.} (\varepsilon_+ \cdot \varepsilon_-)^2 = D(D-3)/2.
\label{GravityExactProduct}
\end{equation}

Together, our result reads
\begin{equation}
\begin{split}
\sum_\textrm{pols.} C &= 
	\mathcal{C}_0^{\mu \nu \alpha \beta} \sum_\textrm{pols.} \varepsilon_{+\mu} \varepsilon_{+\nu} \varepsilon_{-\alpha} \varepsilon_{-\beta} + 
	\mathcal{C}_1^{\mu \nu} \sum_\textrm{pols.} \varepsilon_{+\mu} \varepsilon_{-\nu} (\varepsilon_+ \cdot \varepsilon_-) + 
	\mathcal{C}_2 \sum_\text{pols.} (\varepsilon_+ \cdot \varepsilon_-)^2 \\
&= 
	\mathcal{C}_0^{\mu \nu \alpha \beta} \frac{1}{2}\left(\eta_{\mu \alpha} \eta_{\nu \beta} + 
	\eta_{\mu \beta} \eta_{\nu \alpha} - \frac{2}{D-2}\eta_{\mu \nu}\eta_{\alpha \beta} \right) +
	\mathcal{C}_1^{\mu \nu} \eta_{\mu \nu} \frac{D(D-3)/2}{D-2} +
	\mathcal{C}_2 D(D-3)/2.
\label{GravityRulesResult}
\end{split}
\end{equation}
Eq.~(\ref{GravityRulesResult}) suggests that performing the physical-state sum of the graviton leg in question is equivalent to using the replacement rules
\begin{equation}
\begin{split}
x_1^\mu x_2^\nu x_3^\alpha x_4^\beta \sum_\textrm{pols.} 
\varepsilon_{+\mu} \varepsilon_{+\nu} \varepsilon_{-\alpha} \varepsilon_{-\beta} &\rightarrow x_1^\mu x_2^\nu x_3^\alpha x_4^\beta
\frac{1}{2}\left(\eta_{\mu \alpha} \eta_{\nu \beta} + 
	\eta_{\mu \beta} \eta_{\nu \alpha} - \frac{2}{D-2}\eta_{\mu \nu}\eta_{\alpha \beta} \right),\\
x_1^\mu x_2^\nu \sum_\textrm{pols.} \varepsilon_{+\mu} \varepsilon_{-\nu} (\varepsilon_+ \cdot \varepsilon_-) &\rightarrow
	x_1^\mu x_2^\nu \eta_{\mu \nu} \frac{D(D-3)/2}{D-2},
\end{split}
\end{equation}
along with Eq.~(\ref{GravityExactProduct}) in the individual terms in $\sum_\textrm{pols.} C$. By $x_i$ we denote the various vectors that appear in the calculation.
In this way, we may avoid introducing spurious singularities if we treat the various terms separately.

\subsection{Summary}
Here we summarize our results. We emphasize that we need to reduce the gauge-invariant quantity under study to a basis of Lorentz-invariant products before we use the following replacement rules. We use the right arrow ($\rightarrow$) to indicate that the relation stated is not true term by term; but rather, when we apply it to each term, we reproduce the whole expression correctly. We emphasize that our rules are not equivalent to simply dropping the spurious-singularity pieces. For convenience we repeat Eqs.~(\ref{prYM}) and (\ref{prGR}).

\subsubsection{Gauge theory}

The physical-state projector:
\begin{equation}
P^{\mu \nu} (p,q) = \sum_{\text{pols.}} \varepsilon^\mu (-p) \varepsilon^\nu (p) = \eta^{\mu \nu} - \frac{q^\mu p^\nu + p^\mu q^\nu}{q\cdot p}.
\end{equation}
Equivalent spurious-singularity-free rules:
\begin{equation}
\begin{split}
x_1^\mu x_2^\nu \sum_\textrm{pols.} \varepsilon_\mu (-p) \varepsilon_\nu (p) &\rightarrow x_1^\mu x_2^\nu \eta_{\mu \nu}, \\
 \sum_\textrm{pols.} \varepsilon (-p) \cdot \varepsilon (p) &= D-2,
\end{split}
\label{ymRules}
\end{equation}
where the $x_i$ refer to momenta and polarization vectors that appear in the problem.

\subsubsection{Gravity}
The physical-state projector:
\begin{equation}
P^{\mu \nu \alpha \beta} (p,q) = \sum_{\text{pols.}} \varepsilon^{\mu \nu} (-p) \varepsilon^{\alpha \beta} (p) = \frac{1}{2} \left( 
	P^{\mu \alpha} P^{\nu \beta} + P^{\nu \alpha} P^{\mu \beta} \right) - \frac{1}{D-2} P^{\mu \nu} P^{\alpha \beta},
\end{equation}
where
\begin{equation}
\varepsilon^{\mu \nu} (-p) = \varepsilon^\mu(-p) \varepsilon^\nu(-p), \quad \varepsilon^{\alpha \beta} (p) = 
\varepsilon^\alpha(p) \varepsilon^\beta(p).
\end{equation}
Equivalent spurious-singularity-free rules:
\begin{equation}
\begin{split}
x_1^\mu x_2^\nu x_3^\alpha x_4^\beta \sum_\textrm{pols.} 
\varepsilon_\mu (-p) \varepsilon_\nu (-p) \varepsilon_\alpha (p) \varepsilon_\beta (p) &\rightarrow x_1^\mu x_2^\nu x_3^\alpha x_4^\beta
\frac{1}{2}\left(\eta_{\mu \alpha} \eta_{\nu \beta} + 
	\eta_{\mu \beta} \eta_{\nu \alpha} - \frac{2}{D-2}\eta_{\mu \nu}\eta_{\alpha \beta} \right),\\
x_1^\mu x_2^\nu \sum_\textrm{pols.} \varepsilon_\mu (-p) \varepsilon_\nu (p) \Big( \varepsilon (-p) \cdot \varepsilon (p) \Big) &\rightarrow
	x_1^\mu x_2^\nu \eta_{\mu \nu} \frac{D(D-3)/2}{D-2},\\
\sum_\textrm{pols.} \Big(\varepsilon (-p) \cdot \varepsilon (p) \Big)^2 &= D(D-3)/2, 
\end{split}
\label{grRules}
\raisetag{1.5\normalbaselineskip}
\end{equation}
where the $x_i$ refer to momenta and polarization vectors that appear in the problem.

\subsection{An example in detail\label{spRulExDetSec}}

\begin{figure}[t]
    \centering
    \includegraphics[scale=0.6]{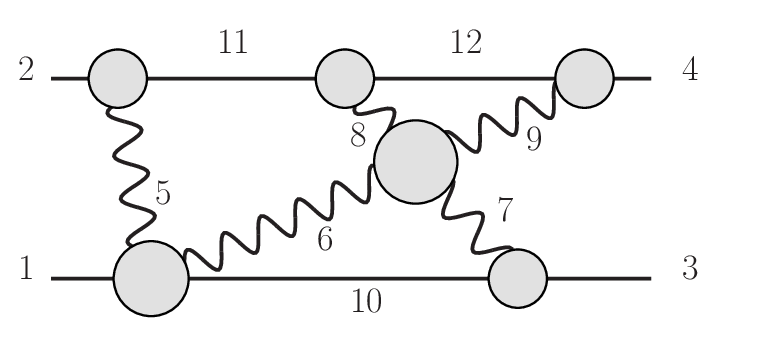}
    \caption{\label{3loopCutFigure}A three-loop generalized-unitarity cut relevant in the construction of the conservative two-body Hamiltonian for spinless black holes to order $G^4$. The blobs represent tree-level amplitudes. The solid lines correspond to scalars while the wiggly ones to gravitons. We take the external particles to be outgoing and the internal momenta to go upwards and to the right. All exposed lines are taken as on shell.}
\end{figure}

In this subsection we discuss the computation of the three-loop generalized-unitarity cut depicted in Fig.~\ref{3loopCutFigure}. This is one of the cuts that can be used to construct the conservative two-body Hamiltonian for spinless black holes to order $G^4$ following the methods of Ref.~\cite{3PM}.

The solid black lines denote massive scalar particles, while the wiggly lines denote gravitons. The blobs correspond to tree-level amplitudes. We may construct them straightforwardly using the Kawai, Lewellen and Tye relations~\cite{Kawai:1985xq} or the Bern, Carrasco and Johansson double copy~\cite{BCJ} from the corresponding gauge-theory ones. The exposed lines are taken as on shell. To construct the cut, we have to sew together the tree-level amplitudes, by summing over the physical states propagating through the internal exposed lines.  

We take all external particles to be outgoing. We take the internal momenta to flow upwards and to the right. The mass of the lower scalar line is $m_1$, while the mass of the upper one is $m_2$. In what follows, it is important to impose the on-shell conditions for both external and internal exposed lines.

We avoid introducing spurious singularities by taking the following steps:
\begin{itemize}
\item We start by sewing particle 5. To reduce our expression to a basis we, for example, solve momentum conservation as $p_{10} = -p_1-p_5-p_6$ and $p_{11} = -p_2-p_5$ and impose the remaining on-shell conditions. We may then use the rules developed above. 
\item We continue to sew leg 6. We must use all on-shell conditions. For example, terms of the form $\varepsilon_i \cdot p_i$ introduced by the above sewing must be set to zero, before applying the replacement rules. 
\item We now sew leg 7. Sewing leg 6 introduced $\varepsilon_i^2$, with $i=7,8,9$, which must be set to zero before the sewing of leg 7. After we impose all on-shell conditions we may use the spurious-singularity-free replacement rules.
\item In a similar manner we sew successively legs 8 and 9.
\end{itemize}

We have explicitly verified that this process gives the correct answer, i.e. the one we get by using the physical-state projectors. However, following our approach we do not have to introduce spurious singularities in any of the steps taken.

\section{Generalized Ward identity\label{GWIsec}}

In this section we develop an alternative approach to simplifying the physical-state sums. In this approach we bring the amplitudes we wish to sew in a form in which they obey the generalized Ward identity (GWI) for the gauge-particle legs of interest. The GWI is a stronger version of the WI, where the vanishing of the amplitude when we replace a polarization vector with the corresponding momentum (Eq.~\ref{WI}) happens without using the special properties of the remaining polarization vectors. Namely, we do not need to use the null~($\varepsilon_i^2 = 0$) and transversallity~($\varepsilon_i \cdot p_i = 0$) conditions.

We organize this section as follows. We start by demonstrating the idea in a simple example. We proceed to describe how we can manipulate any $n$-point amplitude into obeying the GWI for up to $(n-2)$ external gauge particles. If the external gauge particles are $(n-1)$ or less, then all of them may obey the GWI. Next, we explain how this property allows us to drop the spurious-singularity pieces when we insert the physical-state projectors (Eqs.~(\ref{prYM}) and (\ref{prGR})). We discuss some implementation details and limitations of our method. Finally, we conclude by repeating the analysis of the three-loop generalized-unitarity cut of the last section using this second approach.

\subsection{Demonstration in a simple example\label{demoAWIsec}}

We wish to introduce the GWI-based approach in terms of a simple example.  
In contrast to the procedure described in Sect.~\ref{spFreeRules}, which targets the sewing step directly, here we manipulate the amplitude we wish to sew in order to bring it to an appropriate form.
As we discuss in the following subsections, once the gauge-invariant quantity in question satisfies the GWI, we may insert the physical-state projects with the spurious singularities dropped.

Similarly to Sect.~\ref{demoRulesSec}, we demonstrate our method using the Compton amplitude of scalar QED, due to the compactness of the expression. 
We show that this amplitude in a generic form does not obey the GWI. We observe that we may use momentum conservation to bring it to a form such that it does obey the GWI. 
Refs.~\cite{diVecchia,Bern:2020buy} obtained the corresponding gravitational amplitudes in such a form. That was possible due to the simplicity of the four-point amplitude at hand. Our methods, however, are systematic and applicable to any gauge-invariant quantity.

The amplitude is given by
\begin{equation}
\mathcal{A}_4 = 2 i \left( \frac{p_4 \cdot \varepsilon_1 p_3 \cdot \varepsilon_2}{p_1 \cdot p_4} + 
\frac{p_4 \cdot \varepsilon_2 p_3 \cdot \varepsilon_1}{p_1 \cdot p_3} + \varepsilon_1 \cdot \varepsilon_2 \right),
\label{sQEDEq}
\end{equation}
where we set the coupling to unity for convenience. We depict the Feynman diagrams we need in order to calculate this amplitude in Fig.~\ref{sQEDFig}. 
We reduce the amplitude to a basis of Lorentz-invariant products, as we discuss in Sect.~\ref{demoRulesSec}. Specifically, we use momentum conservation to eliminate $p_2$ and $\varepsilon_2 \cdot p_1$ from our basis. 

To discuss the GWI, it is convenient to replace the polarization vectors with generic vectors $w_1$ and $w_2$. We do not assume any special properties for $w_1$ and $w_2$. We have
\begin{equation}
\mathcal{A}_4^w = 2 i \left( \frac{p_4 \cdot w_1 p_3 \cdot w_2}{p_1 \cdot p_4} + 
\frac{p_4 \cdot w_2 p_3 \cdot w_1}{p_1 \cdot p_3} + w_1 \cdot w_2 \right).
\label{comOff1}
\end{equation}
$\mathcal{A}_4^w$ is equivalent to the off-shell amplitude $\mathcal{A}_4^{\mu \nu}$ defined by $\mathcal{A}_4 = \mathcal{A}_4^{\mu \nu} \varepsilon_{1 \mu} \varepsilon_{2 \nu}$. In terms of this object the statements of the WI and the GWI for particle 1 take the form:
\begin{equation}
\mathcal{A}_4^w \big|_{w_1 \rightarrow p_1, w_2 \rightarrow \varepsilon_2} = 0 \quad \text{(WI)}, \quad 
\mathcal{A}_4^w \big|_{w_1 \rightarrow p_1} = 0 \quad \text{(GWI)}.
\end{equation}
We now establish that $\mathcal{A}_4$ in this form does not obey the GWI. Indeed,
\begin{equation}
\mathcal{A}_4^w \big|_{w_1 \rightarrow p_1} = 2i ( p_3 + p_4 + p_1 ) \cdot w_2 = -2i p_2 \cdot w_2 \neq 0.
\end{equation}
The WI however is satisfied, since $\varepsilon_2 \cdot p_2 = 0$.

We draw the reader's attention to the fact that the replacement $w_1 \rightarrow p_1$ introduces $p_1 \cdot w_2$, a Lorentz-invariant product that is not present in our off-shell amplitude (Eq.~(\ref{comOff1})). Once our basis contains all products that may arise from the replacement $w_1 \rightarrow p_1$, then our amplitude will obey the GWI. 

We want to change our basis so that the off-shell amplitude contains $p_1 \cdot w_2$. To do that we use $p_4 = -p_1 - p_2 - p_3$ in our on-shell amplitude. We get
\begin{equation}
\mathcal{A}_4 = 2 i \left( \frac{(p_2+p_3) \cdot \varepsilon_1 p_3 \cdot \varepsilon_2}{p_1 \cdot (p_2+p_3)} - 
\frac{(p_1+p_3) \cdot \varepsilon_2 p_3 \cdot \varepsilon_1}{p_1 \cdot p_3} + \varepsilon_1 \cdot \varepsilon_2 \right),
\end{equation}
which then gives
\begin{equation}
\tilde{\mathcal{A}}_4^w = 2 i \left( \frac{(p_2+p_3) \cdot w_1 p_3 \cdot w_2}{p_1 \cdot (p_2+p_3)} - 
\frac{(p_1+p_3) \cdot w_2 p_3 \cdot w_1}{p_1 \cdot p_3} + w_1 \cdot w_2 \right).
\end{equation}
We introduce the tilde to differentiate this form from the previous one. We may think of $\mathcal{A}_4^w$ and $\tilde{\mathcal{A}}_4^w$ as the off-shell amplitude calculated in different gauge choices. Now, we may confirm that this form satisfies the GWI. Indeed,
\begin{equation}
\mathcal{A}_4^w \big|_{w_1 \rightarrow p_1} = 2i \big( p_3 -(p_1+p_3) + p_1 \big) \cdot w_2 =  0.
\end{equation}

We may verify that the analysis is identical for particle 2. Also, the result would be the same if we were to solve momentum conservation as $p_3 = -p_1-p_2-p_4$.

Let us recap what we just did. We wrote a form of $\mathcal{A}_4$ where $p_1$ and $p_2$ are explicit. We used that form to construct the off-shell amplitude $\tilde{\mathcal{A}}_4^w$. The observation then was that replacing $w_1$ with $p_1$ does not introduce Lorentz-invariant products that were not already in the basis used to write the amplitude. We saw that this property was sufficient for our amplitude to obey the GWI.

We used the generic vectors $w_1$ and $w_2$ in this section for pedantic reasons. Namely, one may use the special properties of the polarization vectors when putting the on-shell amplitude in the appropriate form. However, one should not use them in order to check whether the GWI is satisfied. To alleviate any confusion on that point, we chose to use a different symbol. 

\subsection{The general case}

In this subsection we formulate and prove a claim regarding the GWI in the context of a general amplitude. We start by obtaining a criterion on whether an amplitude satisfies the GWI. Subsequently, using this criterion we obtain a constructive proof of our claim.

We express our amplitude in terms of a basis of Lorentz-invariant products. For brevity we refer to the  Lorentz-invariant products that contain at least one polarization vector as $\varepsilon$-products. Observe that if we perform the mapping $\varepsilon_i \rightarrow p_i$, for a given particle $i$, on the $\varepsilon$-products, then the amplitude vanishes according to the WI (Eq.~(\ref{WI})). To establish whether the amplitude obeys the GWI, we may use the following criterion: 
\begin{itemize}
\item[1.] If, under $\varepsilon_i \rightarrow p_i$ for a given particle $i$, we introduce $\varepsilon$-products that are not part of the basis, then the special properties of the polarization vectors are needed to ensure Eq.~(\ref{WI}). The amplitude does not obey the GWI for particle $i$.
\item[2.] If, under $\varepsilon_i \rightarrow p_i$ for a given particle $i$, all $\varepsilon$-products introduced are part of the basis, then Eq.~(\ref{WI}) is satisfied without the need to use the special properties of the polarization vectors. The amplitude obeys the GWI for particle $i$.
\end{itemize}

At this point we want to comment on the gravitational case, where we have two factors of the polarization vector $\varepsilon_i^\mu$. We use the above criterion on the individual $\varepsilon$-products, rather than the amplitude itself. Each $\varepsilon$-product contains up to one factor of each polarization vector since $\varepsilon_i^2 =0$. Hence, we do not need to modify our criterion in order to use it in gravity. 

\textit{Claim:} An $n$-point amplitude can always be promoted to obey the GWI for up to $(n-2)$ external gauge particles. If the external gauge particles are $(n-1)$ or less, then all of them may obey the GWI. 

\textit{Proof of claim:}
Consider first an $n$-point amplitude where all external particles are gauge particles. To construct our basis, we write down all possible Lorentz-invariant products and we restrict them using the on-shell conditions. First, we remove any products of the form $\varepsilon_i \cdot p_i$ or $\varepsilon_i^2$, due to the polarization vectors being transverse and null. Next, we choose the two particles for which the GWI will not be satisfied. Say we choose particles $n$ and $(n-1)$. We solve momentum conservation in terms of $p_n$. Then, our basis does not include any products of the form $\varepsilon_i \cdot p_n$. Further, again using momentum conservation, we may eliminate $\varepsilon_n \cdot p_{n-1}$. The on-shell conditions put restrictions on momentum products as well, but those are not important for our purposes.

Now, we perform the mapping $\varepsilon_i \rightarrow p_i$ for a given particle $i$ on our basis elements and check whether we introduce $\varepsilon$-products not included in our basis. Namely, we look for the elements $\varepsilon_j \cdot p_n$ for any $j$ or $\varepsilon_n \cdot p_{n-1}$. For particles $n$ and $(n-1)$ we find that we do introduce such elements:
\begin{equation}
\begin{split}
\varepsilon_n \cdot \varepsilon_{n-1} \rightarrow p_n \cdot \varepsilon_{n-1} \quad & \textrm{under} \quad \varepsilon_n \rightarrow p_n, \\
\varepsilon_n \cdot \varepsilon_{n-1} \rightarrow \varepsilon_n \cdot p_{n-1} \quad & \textrm{under} \quad \varepsilon_{n-1} \rightarrow p_{n-1}.
\end{split}
\end{equation}
Performing a similar check for the rest of the particles we find that all elements map to ones within the basis. Hence our amplitude obeys the GWI for the first $(n-2)$ gauge particles.

Finally, consider an $n$-point amplitude that has at least one particle that is not a gauge particle. We label the momenta such that this particle is the $n$-th particle and we solve momentum conservation in terms of $p_n$. Further, since this particle is not a gauge particle, there is no product of the form $\varepsilon_n \cdot p_{n-1}$ to eliminate from our basis. Observe that this was the element blocking the amplitude with $n$ gauge particles to obey the GWI for particle $(n-1)$ in the above setup. Therefore, in this case, the amplitude obeys the GWI for all gauge particles.

\subsection{Simplifying the physical-state projectors using the GWI}

In this subsection we formulate and prove a claim regarding the use of the GWI to drop the spurious singularities when inserting the projectors. We refer to the projectors where we have dropped the spurious-singularity pieces as \textit{simplified projectors}. For simplicity, we look at the gauge-theory case. The discussion in gravity follows in the same way.

\textit{Claim:} Consider the sewing of amplitudes $\mathcal{A}_L$ and $\mathcal{A}_R$. Assuming that $\mathcal{A}_L$ is an $n_L$-point amplitude and $\mathcal{A}_R$ an $n_R$-point one with $n_L \leq n_R$, we may use the WI and the GWI to drop the spurious singularities in up to $(n_L-1)$ insertions of the projector.

\textit{Proof of claim:} We consider the sewing of $n$ gauge particles with $n\leq n_L-1$. We assume that the $n$ particles in question are labeled particle 1 through particle $n$. Using the process described in the previous subsection, we arrange the amplitudes such that they obey the GWI for particles 2 through $n$.

We first show how we may use the WI for particle 1 to drop the spurious singularities in the corresponding insertion of the projector.  We then discuss the use of the GWI for particle 2 for the same purpose. 
Subsequent sewings follow in the same manner.

We sum over the physical polarizations of particle 1, which we denote by $\sum_1$. The amplitude $\mathcal{A}_L$ contains the polarization vector $\varepsilon(p_1) \equiv \varepsilon_1$, while $\mathcal{A}_R$ contains $\varepsilon(-p_1) \equiv \varepsilon_{-1}$. Focusing on the spurious singularity piece of the projector (Eq.~(\ref{prYM})), we have
\begin{equation}
\left( \sum_1 \mathcal{A}_L \mathcal{A}_R \right) _\text{spurious sing.} = 
\frac{1}{p_1 \cdot q} \left( \mathcal{A}_L\big|_{\varepsilon_1 \rightarrow p_1} 
\mathcal{A}_R\big|_{\varepsilon_{-1} \rightarrow q} + \mathcal{A}_L\big|_{\varepsilon_1 \rightarrow q} 
\mathcal{A}_R\big|_{\varepsilon_{-1} \rightarrow p_1} \right) = 0,
\end{equation}
due to the WI (Eq.~(\ref{WI})) of the two amplitudes. Then the complete sum over the physical polarizations of particle 1 gives
\begin{equation}
\sum_1 \mathcal{A}_L \mathcal{A}_R = \big( \mathcal{A}_L \mathcal{A}_R \big)
\big|_{\varepsilon_1^\mu \varepsilon_{-1}^\nu \rightarrow \eta^{\mu \nu}}.
\end{equation}

It is convenient to rewrite the above as follows. We introduce $D$ appropriately normalized vectors $e_i$, such that 
\begin{equation}
\sum_{i=1}^D e_i^\mu e_i^\nu = \eta^{\mu \nu}.
\end{equation}
Then we may write
\begin{equation}
\sum_1 \mathcal{A}_L \mathcal{A}_R = \sum_{i=1}^D \left( \mathcal{A}_L \big|_{\varepsilon_1 \rightarrow e_i}
\mathcal{A}_R\big|_{\varepsilon_{-1} \rightarrow e_i} \right).
\end{equation}

Next, we continue to sum over the physical polarizations of particle 2. We see that the WI of the two amplitudes is not sufficient to guarantee the vanishing of the spurious singularities. Indeed, the WI relies on the special properties of all the polarization vectors. Since the first pair of polarization vectors is replaced by the vectors $e_i$, the WI for the two amplitudes is not obeyed.

At this point we want to emphasize that the entire expression still obeys the WI for the remaining gauge particles. However, it is the WI of the individual amplitudes, $\mathcal{A}_L$ and $\mathcal{A}_R$, that makes the spurious-singularity pieces vanish.

We may now see why the GWI is helpful. Since it does not rely on the special properties of the polarization vectors, it still holds after we replace them with the vectors $e_i$. For example, if $\mathcal{A}_L$ satisfies the GWI for particle 2, we have
\begin{equation}
\mathcal{A}_L \big|_{\varepsilon_1 \rightarrow e_i, \varepsilon_2 \rightarrow p_2} = 0,
\end{equation}
for any $i$.  Therefore, as long as both amplitudes obey the GWI for the particle being sewn, we may drop the spurious-singularity pieces from the insertion of the projector.  This concludes our proof.

We now comment on a limitation of this approach. If in the above setup we wish to sew all $n_L$  particles, as is the case in some calculations of matrix-element squares for example, then we cannot avoid introducing spurious singularities in the last insertion of the projector. 
The situation is different however if the $n_L$ sewings involve three or more amplitudes. We take up this discussion in the next subsection.

\subsection{Implementation details}

\begin{figure}[t]
	\includegraphics[scale=.55]{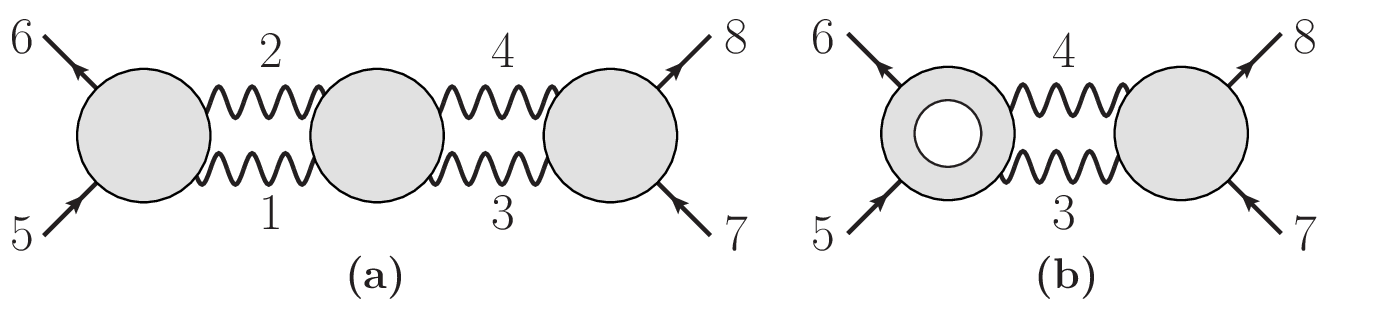}
    \caption{\label{Hcut} \small An example where we need to break the process into two steps. All exposed lines are taken as on-shell. Solid blobs represent tree-level amplitudes. In (b) the hollow blob represents the one-loop quantity we get by performing the physical-state sums for particles 1 and 2 on the two amplitudes on the left in (a).}    
\end{figure}

In this subsection we comment on some details that are important for the implementation of the method developed above. We start by discussing how can we approach the sewing of three or more amplitudes, as for example is shown in Fig.~\ref{Hcut}. Then, we comment on when exactly should we use the special properties of the polarization vectors.

In a multiloop calculation based on generalized-unitarity we typically need to sew together multiple amplitudes. To maximize the efficiency of our method in this case we should break the process in \textit{steps}. In any given step we should be sewing together exactly two amplitudes. We recall that by amplitude we refer to any gauge-invariant quantity. 

Take for example the case depicted in Fig.~\ref{Hcut}a. To build the two-loop generalized-unitarity cut in question we need to sew together three tree-level amplitudes. Given that we need to sew all legs of the middle four-point amplitude, one might think that we are forced to introduce spurious singularities in one of the insertions of the physical-state projector. We circumvent that in the following way.

We start by sewing the two tree-level amplitudes on the left. We prepare them so that the legs that carry the momenta $\pm p_1$ and $\pm p_2$ obey the GWI. Then we proceed with the sewing using the simplified projectors. Next, we have to sew the one-loop quantity on the left with the tree-level amplitude on the right (Fig.~\ref{Hcut}b). Given that both of these objects are gauge-invariant quantities, we may again use the above method. Namely, we prepare  them so that the legs that carry the momenta $\pm p_3$ and $\pm p_4$ obey the GWI. We do this by a simple change of basis, just like before. Then, we proceed with the sewing using the simplified projectors. In this way we may sew multiple amplitudes together, introducing the minimum number of spurious singularities.

An aspect that this method hinges on is the appropriate use of the special properties of polarization vectors. Specifically, we first manipulate the amplitudes into the appropriate form as discussed above. In doing so we should use all special properties of the polarization vectors. Then, we sew a number of legs of the amplitudes. Between these sewings we should not use any special properties of the polarization vectors, as that would interfere with the GWI of the legs left to be sewn. After all sewings of that step are completed, we should again use the special properties of the polarization vectors. 

For example, referring again to Fig.~\ref{Hcut}, upon sewing particle 1, we introduce $\varepsilon_2 \cdot p_2$ and  $\varepsilon_3 \cdot p_3$ among other terms. We should not set these terms to zero before sewing particle 2, as that would interfere with the GWI for that particle. After we complete the sewings of the left two amplitudes (i.e. the sewings of particles 1 and 2), we change our basis to make our amplitudes obey the GWI for particles 3 and 4. We should now use the special properties of the remaining polarization vectors. In our example, we should set $\varepsilon_3 \cdot p_3$ to zero, among other terms.

\subsection{Summary}

Here we briefly summarize the method developed above. The problem at hand is to organize a calculation in gauge theory or gravity where multiple $D$-dimensional sewings are required, as is for example typical for generalized-unitarity approaches to multiloop calculations. The conventional way of performing such a calculation is to use the physical-state projectors,  Eqs.~(\ref{prYM}) and (\ref{prGR}), repeated here for convenience, 
\begin{equation}
P^{\mu \nu} (p,q) = \sum_{\text{pols.}} \varepsilon^\mu (-p) \varepsilon^\nu (p) = \eta^{\mu \nu} - \frac{q^\mu p^\nu + p^\mu q^\nu}{q\cdot p},
\end{equation}
and
\begin{equation}
P^{\mu \nu \alpha \beta} (p,q) = \sum_{\text{pols.}} \varepsilon^{\mu \nu} (-p) \varepsilon^{\alpha \beta} (p) = \frac{1}{2} \left( 
	P^{\mu \alpha} P^{\nu \beta} + P^{\nu \alpha} P^{\mu \beta} \right) - \frac{1}{D-2} P^{\mu \nu} P^{\alpha \beta} .
\end{equation}

Our proposed method allows one to perform the sewings without introducing spurious singularities. Specifically, we bring the amplitudes to be sewn in a form such that we may replace $P^{\mu \nu} (p,q)$ with $\eta^{\mu \nu}$ in the above two equations. We refer to these projectors as simplified projectors.

To do so we first need to organize our calculation in steps. In each step we are only sewing two gauge-invariant quantities together. Before the sewing, we use momentum conservation and the on-shell conditions so that the momenta of the particles to be sewn are some of the independent momenta appearing in our expression. We may then verify that the amplitudes at hand obey the GWI for these legs. We proceed with the sewing using the simplified projectors. During a given step, between sewings, it is important not to use the on-shell conditions, as that spoils the GWI for the remaining legs to be sewn. Finally, after all sewings of a given step are completed, we may again use the on-shell conditions. 
If in any given step we have to sew all external particles of an amplitude and those particles are gauge-particles, then we have to introduce spurious singularities in one of the insertions of the projector.

\subsection{An example in detail\label{AWIexDetSec}}

Here we repeat the discussion of the construction of the three-loop generalized-unitarity cut depicted in Fig.~\ref{3loopCutFigure} using the GWI approach.
To construct the cut of Fig.~\ref{3loopCutFigure} without introducing any spurious singularities we may follow these steps:
\begin{itemize}
\item We start by sewing particle 5. We need to verify that the associated three-point and four-point tree-level amplitudes satisfy the GWI for that particle. To do so, we choose to use momentum conservation to eliminate momenta 10 and 11. Next, we impose the on-shell conditions. Now our amplitudes obey the GWI for particle 5 and we may proceed to sew it using the simplified projector.
\item Next, we choose to sew particle 10. Since that particle is scalar, this simply amounts to multiplying the corresponding three-point amplitude with the result we got above.
\item We continue by sewing particles 6 and 7. To do this we must manipulate both the expression we got so far and the four-point tree-level amplitude to obey the GWI for particles 6 and 7. 
\begin{itemize}
\item For the four-point amplitude we choose to solve momentum conservation as $p_9 = p_6+p_7-p_8$. Further, we impose $\varepsilon_9 \cdot p_8 = \varepsilon_9 \cdot p_7 + \varepsilon_9 \cdot p_6$ along with the remaining on-shell conditions. In this way we choose a basis where the momenta $p_6$ and $p_7$ are independent, turning the amplitude into obeying the GWI for the corresponding legs.
\item For the expression we got from the previous sewings, we use $p_3 = -p_1-p_2-p_4-p_6-p_7$. We impose the on-shell conditions for all exposed lines, even the ones already sewn (e.g. $p_{10}^2 -m_1^2 = p_5^2 = 0$).
\end{itemize}  
We may now sew particles 6 and 7 using the simplified projector. It is important not to use any on-shell conditions between the two sewings. If we do, after sewing particle 6 for example, we spoil the GWI for particle 7. Namely, sewing particle 6 introduces $\varepsilon_7 \cdot p_7$ and $\varepsilon_7^2$. We do not set these terms to zero at this point, as they are essential in reproducing the correct result.
\item We may now proceed to sew leg 8 and then leg 9, following a similar approach. Alternatively, we may multiply the remaining three-point amplitudes together to build a four-point quantity, which we then sew to the expression we got so far. We choose to do the latter in order to emphasize that we may sew any gauge-invariant quantity with the proposed method. Again, we should make sure that the two quantities obey the GWI for particles 8 and 9.
\begin{itemize}
\item For the four-point quantity, we choose to solve momentum conservation as $p_{11} = p_8+p_9-p_3$. 
\item For the expression we got so far we use $p_{11} = -p_1-p_2-p_4-p_8-p_9$. We should also choose one of the $p_6$, $p_7$ and $p_{10}$ as our loop momentum and solve for the other two in terms of it. Then we impose the on-shell conditions. A new feature that appears here is the introduction of $\varepsilon_8^2$ and $\varepsilon_9^2$ from sewing particles 6 and 7. At this stage we should use the null condition and set them to zero. 
\end{itemize}
After these manipulations we may sew the final two legs using the simplified projectors.
\end{itemize}

We comment again on the subtle point that we should not use the special properties of the polarization vectors between sewings of a given step, but we have to use them in order to make a quantity obey the GWI. This is why we do not set $\varepsilon_7^2$ to zero after sewing particle 6 but we do set $\varepsilon_8^2$ and $\varepsilon_9^2$ to zero after sewing particles 6 and 7.

We have explicitly verified that this process correctly reproduces the answer we get by using the full projector. We emphasize once more that we do not have to introduce spurious singularities at any of the steps taken.

\section{Conclusions\label{ConclusionsSec}}	

$D$-dimensional approaches are useful in a variety of
problems. Specifically, they are natural in the context of dimensional
regularization~\cite{Collins}. In this setup it is often the case that we
need to perform a physical-state sum of a gluon or graviton leg. An important example where such sums appear is calculations based on
$D$-dimensional generalized unitarity~\cite{Unitarity,dDimUnitarity}.  Other
examples include $D$-dimensional
matrix-element
squares for cross sections~\cite{scatteringCrossSections}, and the decomposition of an amplitude into
gauge-invariant tensors \cite{gaugeInvariantTensors}.  

In
general, these physical-state sums introduce spurious
light-cone singularities. These spurious singularities unnecessarily
complicate the expressions, especially in the gravitational
case. Further, if we do not eliminate them prior to phase-space or loop
integration, the spurious singularities require nontrivial
prescriptions to make them well defined. Hence, with new generalized-unitarity based calculations ~\cite{3PM,Bern:2020buy} pushing the frontier of relativistic gravitational-wave physics, there is a specific need for methods that avoid introducing these spurious singularities.

In this paper we achieve this goal by providing two independent
methods that allow us to perform the physical-state sums so that we do not
introduce spurious singularities. Our methods are applicable to any
gauge-invariant quantities, at tree or loop level.
In our first method we identify gauge-invariant subpieces in our
expression. We observe that we may perform the physical-state sum for
each piece without introducing spurious singularities. In this way we
derive a set of replacement rules that do not contain spurious singularities and are equivalent to the physical-state
projectors.
In our second method we use momentum conservation to bring our
gauge-invariant quantities in a form such that they obey the generalized Ward identity for the legs
to be sewn. In this form, the spurious singularities automatically
drop out of the physical-state sums. We identify certain limitations on the applicability of this approach. 

Spurious singularities also appear in physical-state sums of massive vector bosons, spin-3/2 particles, or other higher-spin fields. We are confident that we can tackle these problems with methods similar to the ones developed in this paper.
We hope that our methods will
help simplify future calculations involving $D$-dimensional physical-state sums in gauge theory and gravity.

\begin{acknowledgments}
We thank Chia-Hsien Shen and Andr\'es Luna for fruitful conversations. We especially wish to thank Zvi Bern for many helpful discussions and comments on the manuscript. This work is supported by the U.S. Department of Energy (DOE) under award number DE-SC0009937. We are also grateful to the Mani L. Bhaumik Institute for Theoretical Physics for additional support.
\end{acknowledgments}


\end{document}